\documentclass[twocolumn,aps,prl,showpacs,superscriptaddress,amsfonts,amssymb,amsmath]{revtex4}
\usepackage{graphicx}
\usepackage{graphics}
\begin{document}

\title{Exact solutions of an $SO(5)$-invariant spin-$3/2$ Fermi gas model}

\author{Yuzhu Jiang}
\affiliation{Beijing National Laboratory for Condensed Matter
Physics, Institute of Physics, Chinese Academy of Sciences, Beijing
100190, People's Republic of China}
\author{Junpeng Cao}
\affiliation{Beijing National Laboratory for Condensed Matter
Physics, Institute of Physics, Chinese Academy of Sciences, Beijing
100190, People's Republic of China}
\author{Yupeng Wang*}
\affiliation{Beijing National Laboratory for Condensed Matter
Physics, Institute of Physics, Chinese Academy of Sciences, Beijing
100190, People's Republic of China}

\begin{abstract}
An exactly solvable model describing the dilute spin-$3/2$ fermion
gas in one-dimensional optical trap is proposed. The diagonalization
of the model Hamiltonian is derived by means of the Bethe ansatz
method. Exotic spin excitations such as the heavy spinon with
fractional spin $3/2$, the neutral spinon with spin zero and the
dressed spinon with spin $1/2$ are found based on the exact
solution.
\end{abstract}


%05.30.Fk   Fermion systems and electron gas
%75.10.-b   General theory and models of magnetic ordering
%03.75.Ss   Degenerate fermi gas

\maketitle

The study on the optically and magnetically trapped ultracold atoms
has attracted a lot of attentions in the recent years. By using the
magnetic fields or laser beams, the atoms can be trapped and cooled
down to very low temperatures
\cite{MGreiner01,AGolitz01,FSchreck01,HMoritz03,
h,b1,TKinoshita04,BParedes04,HMoritz05,SAubin05}. A fascinating fact
is that not only the components but also the physical parameters of
such systems can be manipulated in experiments, which allow us to
mimic the ordinary correlated electron systems and even to explore
new states of matter. For examples, by means of the Feshbach
resonance techniques, one can tune the scattering lengths or the
interactions among the atoms; putting an optical lattice on one can
study the behaviors of the atoms in a tunable periodic potential and
with strong anisotropic traps, the one- and two-dimensional quantum
systems can be realized. One of the hot research topics in this
field is the study on the high spin cold atom systems. The physics
of bosonic ultracold atoms with hyperfine spin $F=1$ have been
extensively studied \cite{sbec2,sbec3,mfpt1,e1,e2,cao}. The
degenerate Fermi gas with hyperfine spin $F=3/2$ could be obtained
by cooling alkali atoms $^{132}$Cs, as well as alkaline-earth atoms
$^9$Be, $^{135}$Ba and $^{137}$Ba in experiments. The ground states
of these high spin fermionic atoms in optical traps are very rich.
Many peculiar quantum orders and exotic collective excitations
appear in these systems, which are rare in the ordinary interacting
electron systems \cite{10}. Several methods have been developed to
approach these ultracold atom systems. For examples, Wu {\it et al.}
showed that the one-dimensional (1D) spin-$3/2$ Fermi gas with
s-wave scattering possesses the $SO(5)$-invariance and obtained the
phase diagram of the system by means of bosonization \cite{wu1,wu3};
the functional integral approach was applied on the interacting
spin-$3/2$ fermionic ultracold atoms in 2D square optical lattice to
study a variety of Mott insulating phases \cite{zhang1,zhang2}; two
different superfluid phases in $F=N-1/2$ fermionic cold atom system
were found, i.e., an unconfined BCS pairing phase and a confined
molecular-superfluid made of $2N$ fermions, depending on whether a
discrete symmetry is spontaneously broken or not \cite{p}.

Despite the fast progress in this field, the understanding to the
high-spin correlated systems is still far from satisfied. Reference
models with exact solutions are undoubtedly needed to get deep
insight about these interesting quantum many-body systems. An
important progress in this aspect is made by Controzzi1 and Tsvelik,
who constructed an exactly solvable model for isospin $F=3/2$
fermionic system with linear dispersion relations \cite{c}. In this
Letter, we propose a new exactly solvable model which describes
properly a 1D spin-$3/2$ cold atom system. With the Bethe ansatz
solutions, we derive out exactly the ground state. Exotic spin
excitations such as the heavy spinons with fractional spin $3/2$,
the neutral spinons with zero spin and the dressed spinons with spin
$1/2$ are found.

The fermions with hyperfine spin $F=3/2$ in a 1D trap is
appropriately described by the following model Hamiltonian
\begin{eqnarray}
H = \sum_{j=1}^N \left[-\frac{\partial^2 }{\partial x_j^2}+
V_e(x_j)\right]+ \sum_{l<j, m}^N g_m\delta(x_j-x_l) P^m_{jl},\label{h}
\end{eqnarray}
where $ V_e(x_j)$ is the external potential, $g_m$ is the two-body
coupling constant in the total spin $m=0, 2$ channel and $P_{jl}^m$
is the projection operator onto the spin $m$ channel. Because of the
antisymmetric wave functions of the fermions, non-trivial scattering
processes occur only in the $m=0$ and $m=2$ channels, and those in
the $m=1$ and $m=3$ channels are irrelevant for the
$\delta$-function interaction potential. In the low density case, we
omit the external field. The Hamiltonian (\ref{h}) can be rewritten
as
\begin{eqnarray}
H=-\sum_{j=1}^N \frac{\partial^2}{\partial x_j^2} +
\sum_{l<j}^N(c_0+ c_2{\bf S}_j \cdot {\bf S}_l)\delta(x_j-x_l),
\label{h0}
\end{eqnarray}
where $c_0=g_0/3+2g_2/3$, $c_2=g_2/3-g_0/3$ and ${\bf S}_j$ is the
spin-$3/2$ operator. If $c_2=0$, the system (\ref{h0}) is
$SU(4)$-invariant and was solved exactly by Sutherland \cite{suth}.
In the present Letter, we show that the system (\ref{h0}) possesses
another integrable line  $c_0=c/2$, $c_2=-2c/3$ at which the
physical properties are quite different to those of the Sutherland
model \cite{suth, sch}. An obvious fact is that in our case, the
particle number of an individual spin component is no longer
conserved because of the broken $SU(4)$-symmetry. Nevertheless, the
system (\ref{h0}) is still $SO(5)$-invariant at this new integrable
line, and the following three independent conserved quantities are
hold:
\begin{eqnarray}
&&I_1=N_{3/2}+N_{1/2}+N_{-1/2}+N_{-3/2}, \nonumber\\
&&I_2=N_{3/2}-N_{-3/2}, \\
&&I_3=N_{1/2}-N_{-1/2},\nonumber
\end{eqnarray}
where $N_{s}$ indicates the particle number with the spin component
$s=\pm 1/2, \pm 3/2$.

Assume the wave function takes the following form
\begin{eqnarray}
&&\Psi(x_1s_1, \cdots,
x_Ns_N)=\sum_{Q,P}\theta(x_{Q_1}<\cdots<x_{Q_N})\nonumber \\
&& \quad\quad\quad\quad \times A_{s_1\cdots s_N} (Q,
P)e^{i\sum_{l=1}^{N}k_{P_l}x_{Q_l}},
\end{eqnarray}
where $Q=(Q_1, \cdots, Q_N)$ and $P=(P_1, \cdots, P_N)$ are the
permutations of the integers $1,\cdots, N$; $k_j$ are the quasi
momenta carried by the particles; $\theta(x_{Q_1}<\cdots<x_{Q_N})=
\theta(x_{Q_N}-x_{Q_{N-1}}) \cdots \theta(x_{Q_2}-x_{Q_1})$ and
$\theta(x-y)$ is the step function. With the standard coordinate
Bethe ansatz method, we obtain the two-body scattering matrix as
\begin{eqnarray}
S_{jl}=\frac{k_j-k_l-i\frac{3c}{2}}{k_j-k_l+i\frac{3c}{2}}P_{jl}^0
+P_{jl}^1 +\frac{k_j-k_l-i\frac{c}{2}}{k_j-k_l+i\frac{c}{2}}P_{jl}^2
+ P_{jl}^3,\label{sca}
\end{eqnarray}
which satisfies the Yang-Baxter equation
\begin{eqnarray}
&&S_{12}(k_1-k_2)S_{13}(k_1-k_3)S_{23}(k_2-k_3)\nonumber \\
&&\quad = S_{23}(k_2-k_3)S_{13}(k_1-k_3)S_{12}(k_1-k_2). \label{ybb}
\end{eqnarray}
Applying further the nested algebraic Bethe ansatz with periodic
boundary conditions \cite{Mar}, we obtain the following Bethe ansatz
equations(BAE):
\begin{eqnarray}
&& e^{ik_jL} =
\prod_{\alpha=1}^{M_1}\frac{k_j-\lambda_{\alpha}+i\frac{c}{4}}
{k_j-\lambda_{\alpha}-i\frac{c}{4}}, \quad j=1, \cdots, N, \nonumber\label{bae1}\\
&& \prod_{l=1}^N\frac{\lambda_{\beta}-k_l+i\frac{c}{4}}
{\lambda_{\beta}-k_l-i\frac{c}{4}}
\prod_{\gamma=1}^{M_2}\frac{\lambda_{\beta}-\mu_{\gamma}+i\frac{c}{2}}
{\lambda_{\beta}-\mu_{\gamma}-i\frac{c}{2}} \nonumber \\
&&\quad
=-\prod_{\alpha=1}^{M_1}\frac{\lambda_{\beta}-\lambda_{\alpha}+i\frac{c}{2}}
{\lambda_{\beta}-\lambda_{\alpha}-i\frac{c}{2}}, \quad \beta=1,\cdots, M_1, \label{bae2}\\
&&\prod_{\alpha=1}^{M_1}\frac{\mu_{\nu}-\lambda_{\alpha}+i\frac{c}{2}}
{\mu_{\nu}-\lambda_{\alpha}-i\frac{c}{2}}
=-\prod_{\gamma=1}^{M_2}\frac{\mu_{\nu}-\mu_{\gamma}+ic}
{\mu_{\nu}-\mu_{\gamma}-ic}, \nonumber \\
&&\hspace{4.2cm}  \nu=1,\cdots, M_2, \nonumber \label{bae3}
\end{eqnarray}
where $N=N_{3/2}+N_{1/2}+N_{-1/2}+N_{-3/2}$, $M_1=N_{1/2}
+2N_{-3/2}+N_{-1/2}$, $M_2=N_{-1/2} +N_{-3/2}$ and $L$ is the length
of the system. The corresponding eigen energy of the Hamiltonian is
$E=\sum_{j=1}^N k_j^2$.

We consider the $c>0$, i.e., the repulsive interaction case. By
carefully checking the structure of the BAE (\ref{bae2}), we find
that all the charge rapidities $k_j$ take real values, indicating
the absence of charge bound state. However, the spin rapidities
$\lambda_\beta$ and $\mu_\nu$ may form strings with the following
form in the thermodynamic limit
\begin{eqnarray}
&& \lambda^{n}_{\beta,j} = \lambda^{(n)}_\beta +
\frac{ic}{4}(n+1-2j), \; j=1,\cdots,n, \\
&& \mu^{m}_{\nu,j} = \mu^{(m)}_\nu + \frac{ic}{2}(n+1-2j),\;
j=1,\cdots,m, \label{string_h}
\end{eqnarray}
where $\lambda^{(n)}_\beta$ and $\mu^{(m)}_\nu$ are the real parts
of the $n$-string of $\lambda$ and the $m$-string of $\mu$,
respectively. Denote $\rho$, $\eta_n$ and $\sigma_m$ as the
densities of $k$, $\lambda$ $n$-strings and $\mu$ $m$-strings in the
thermodynamic limit $N\to\infty$, $L\to\infty$ and $N/L\to finite$,
respectively, and $\rho^{h}$, $\eta_n^h$ and $\sigma_m^{h}$ the
corresponding densities of holes. At temperature $T$, the Gibbs free
energy of the system with an external magnetic field $h$ and
chemical potential $A$ is
\begin{eqnarray}
F=E- A N -h\left(\frac{3}{2}N-M_1-M_2\right)-TS,
\end{eqnarray}
where $E=L \int k^2 \rho(k) dk$ is the energy; $N=L \int \rho(k)dk$
is the particle number; $M_1= L\sum_{n} n \int \eta_n(\lambda)d
\lambda$; $M_2=L\sum_{m} m \int \sigma_m(\mu)d \mu$; $S= L\int
\{(\rho+\rho^h)\ln(\rho+\rho^h)-\rho\ln\rho-\rho^h\ln\rho^h
+\sum_{n}[(\eta_n+\eta_n^h)\ln(\eta_n+\eta_n^h) -\eta_n\ln\eta_n
-\eta_n^h\ln\eta_n^h] +
\sum_{m}[(\sigma_m+\sigma_m^h)\ln(\sigma_m+\sigma_m^h)
-\sigma_m\ln\sigma_m-\sigma_m^h\ln\sigma_m^h] \}dk$ denotes the
entropy. Minimizing the Gibbs free energy at the thermal
equilibrium, we obtain the following thermodynamic Bethe ansatz
equations (TBAE)
\begin{eqnarray}
&& \ln \tilde \rho = \frac{k^2-A}{T} -\sum_n a_n*\ln(1+ {\tilde
\eta}_{n}^{-1}),
\nonumber\\
&& \ln {\tilde \eta}_1 = G_1*\ln [(1+\tilde \eta_2)(1+ \tilde
\rho^{-1})^{-1}], \nonumber\\
&& \ln \tilde \sigma_1 = G_2*\ln  \frac{1+ \tilde \sigma_2}
{(1+\tilde \eta_{1}^{-1}) (1+\tilde \eta_{3}^{-1})} \nonumber\\
&&\quad\quad\quad -\frac{G_2}{G_1}* \ln(1+\tilde \eta_2^{-1}),
\nonumber\\
&& \ln \tilde \eta_{n \in even} = G_1*\ln  \frac{(1+ \tilde
\eta_{n-1})
(1+\tilde \eta_{n+1})}{1+\tilde \sigma_{n/2}^{-1}}, \label{bae-3str-1}\\
&& \ln \tilde \eta_{n \in odd} = G_1*\ln[ (1+\tilde \eta_{n-1})
(1+\tilde \eta_{n+1}) ], \nonumber \\
&& \ln \tilde \sigma_m= G_2*\ln  \frac{ (1+\tilde \sigma_{m-1})
(1+\tilde \sigma_{m+1})} {(1+\tilde \eta_{2m-1}^{-1})
(1+\tilde \eta_{2m+1}^{-1})} \nonumber\\
&& \quad\quad\quad -\frac{G_2}{G_1}*\ln(1+\tilde \eta_{2m}^{-1}),
\nonumber \\
&&\lim_{n \rightarrow \infty}\frac{\ln \tilde
\eta_n}{n}=\frac{h}{T}, \quad \lim_{m \rightarrow
\infty}\frac{\ln\tilde \sigma_m}{m}=\frac{h}{T}, \nonumber
\end{eqnarray}
where $\tilde \rho=\rho_h/\rho$, $\tilde \eta_n= \eta_n^h/\eta_n$,
$\tilde \sigma_m =\sigma_m^h/\sigma_m$, $G_n= a_n/(a_0 +a_{2n})$,
$a_0\equiv \delta(x)$, $a_n(x)=4nc/[\pi(16 x^2 + n^2c^2)]$ and
$a_n*f(x) = \int a_n(x-y) f(y) dy$.

The ground state configuration of the system can be obtained by
taking the limit of $T\to 0$ and $h\to 0$. In this case, most of the
string densities are zero and the TBAE (\ref{bae-3str-1}) are
reduced to
\begin{eqnarray}
&&\rho(k)=\frac{1}{2\pi}+ a_1*\eta_1(k)+ a_2*\eta_2(k), \nonumber\\
&& \eta_1(\lambda)=\int_{-Q}^{Q} a_1(\lambda-k) \rho(k) dk -
a_2*\eta_1(\lambda)\nonumber \\
&&\qquad  - (a_1+a_3)*\eta_2(\lambda)  +
a_1*\sigma(\lambda), \label{bae-str-1} \\
&& \eta_2(\xi)=\int_{-Q}^{Q} a_2(\xi-k) \rho(k) dk
-(a_1+a_3)*\eta_1(\xi) \nonumber \\
&&\qquad - (2a_2+a_4)*\eta_2(\xi)  +
(a_1+a_3)*\sigma(\xi), \nonumber \\
&& \sigma(\mu)= a_1*\eta_1(\mu) + (a_1+a_3)*\eta_2(\mu) -
a_4*\sigma(\mu). \nonumber
\end{eqnarray}
The Fermi point $Q$ is determined by the density of particles $N/L =
\int_{-Q}^{Q} \rho(k) dk$. $M_1/L=\int [\eta_1(\lambda) + 2 \eta_2
(\lambda)]d \lambda$ and $M_2/L=\int \sigma(\mu) d\mu$. Such a
ground state configuration is quite different from that of the
$SU(4)$ Sutherland model, where there is no string or spin bound
state in the ground state. In the present $SO(5)$ case, part of the
spin rapidities form 2-strings which heavily affect the spin
excitations as we shall show below. From the solutions of
Eq.(\ref{bae-str-1}), we can easily derive out $M_1=N$ and
$M_2=N/2$, which give the total spin of the ground state $S=3N/2 -
M_1-M_2= 0$, indicating a spin singlet ground state. When $c$ tends
to zero, the function $a_n(x)\to \delta(x)$. In this case we recover
the free Fermi gas solutions
\begin{equation}
\rho(k) = \frac{2}{\pi}, \quad \frac{E}{L} = \frac{4}{3\pi}Q^3,
\quad \frac{N}{L} = \frac{4}{\pi}Q.
\end{equation}
When $c\to \infty$, $a_n(x)\to 0$, we recover the Tonks-Girardeau
solutions
\begin{equation}
\rho(k) = \frac{1}{2\pi}, \quad \frac{E}{L} = \frac{1}{3\pi}Q^3,
\quad \frac{N}{L} = \frac{1}{\pi}Q.
\end{equation}

\begin{figure}[ht]
\begin{center}
\includegraphics[height=6cm,width=8cm]{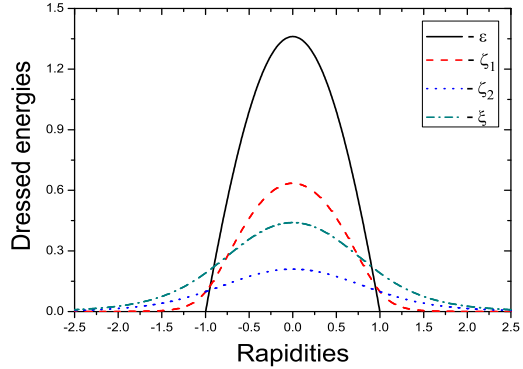}
\end{center}
\caption{(Color online) The dressed energies of the ground state
($Q=1,\,c=1$). $\epsilon$, $\zeta_1$, $\zeta_2$ and $\xi$ are the
dressed energies of rapidities $k$, real $\lambda$, 2-string
$\lambda$ and real $\mu$, respectively. At the Fermi point $Q$, the
dressed energy $\epsilon(Q)$ is zero.} \label{fig1}
\end{figure}

Based on the ground state configuration, the elementary excitations
of the system can be studied exactly. In the integrable models, the
excitation energies are uniquely defined by the so-called dressed
energies $\epsilon = T \ln \tilde \rho$, $\zeta_n = T \ln \tilde
\eta_n$ and $\xi_m=  T \ln \tilde \sigma_m$. In our case, when $T\to
0$, only $\epsilon$, $\zeta_{1,2}$ and $\xi_1$ are left. The dressed
energies of the ground state for $Q=1$ and $c=1$ are shown in
Fig.\ref{fig1}.
\begin{figure}[ht]
\begin{center}
\includegraphics[height=6cm,width=8cm]{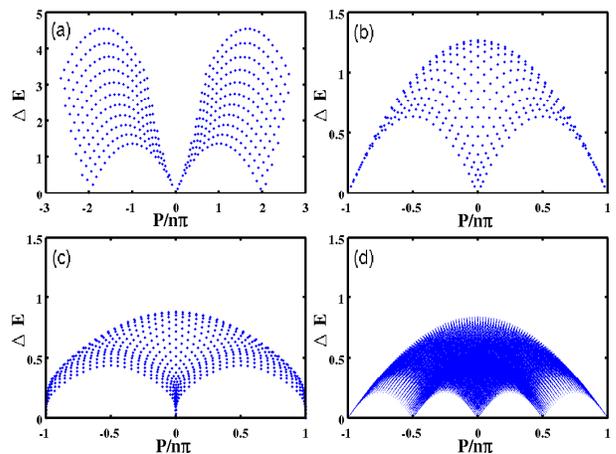}
\end{center}
\caption{(Color online) The low-lying excitations ($c=1$, $Q=1$).
$\Delta E$ and $P$ are the energy and the momentum carried by the
excitation. (a) The charge-hole excitations. (b) The excitation of
two holes of real $\lambda$. (c) The excitation of two holes of real
$\mu$. (d) the excitation of four holes of $\lambda$ 2-string.}
\label{fig2}
\end{figure}
The low-lying excitations of the system can be studied systemically
by adding particles, holes or strings into the ground state
configuration of the rapidities. The excitation energy reads
\begin{eqnarray}
\sl{\Delta
E}&=&\sum_{j=1}^{n_h}\epsilon(k_j^h)+\sum_{j=1}^{n_p}\epsilon(k_j^p)
+\sum_{j=1}^{m_1}\zeta_1(\lambda^h_{1,j})\nonumber\\
&+&\sum_{j=1}^{m_2}\zeta_2(\lambda^h_{2,j})+\sum_{j=1}^{m_3}
\eta_1(\mu^h_j),\label{de}
\end{eqnarray}
where $n_h$, $n_p$ and $m_{1,2,3}$ are the numbers of the charge
holes, excited charges, holes in real $\lambda$ sea, in 2-string
$\lambda$ sea and in real $\mu$ sea, respectively; $k_j^h$, $k_j^p$,
$\lambda^h_{1,2,j}$ and $\mu^h_j$ are the positions of the
corresponding charges and holes. Formally, the extra strings
contribute nothing to the energy because the contribution of such
strings is exactly canceled by the rearrangement of the Fermi sea.
Some of the low-lying excitations are shown in Fig.\ref{fig2}. The
real $k$ charge-hole excitation in our case is similar to that of
$SU(4)$-invariant Sutherland model as shown in Fig.\ref{fig2}(a).
However, the excitations in the spin sector are quite different from
those of the Sutherland model. The spin quanta carried by the spin
excitations reads
\begin{equation}
S=\frac{3}{2}m_2+2m_3+\sum_{l\geq3}(2-l)m_{\lambda^{(l)}} +
\sum_{t\geq2}(1-t)m_{\mu^{(t)}},
\end{equation}
where $m_{\lambda^{(l)}}$ and $m_{\mu^{(t)}}$ are the numbers of
$\lambda$ $l$-strings and the $\mu$ $t$-strings formed in the
excitations, respectively. We note the numbers of holes and strings
added are not independent but satisfy some constraints determined by
the BAE:
\begin{eqnarray}
&&\Delta M_1^1=-m_1+\frac{1}{2}m_2, \nonumber\\{}&&\Delta M_1^2 =
\frac{1}{2}m_1-\frac{3}{4}m_2-\frac{1}{2}m_3 -\sum_{l\geq3}
m_{\lambda^{(l)}}, \\{} &&\Delta M_2^1=-\frac{1}{2}m_2 -m_3
-\sum_{t\geq2} m_{\mu^{(t)}},\nonumber
\end{eqnarray}
where $\Delta M_{1,2}^l$ (integers) indicate the number changes of
$\lambda, \mu$ $l$-strings. Several possible hole configurations are
listed in Table \ref{table}.
\begin{figure}[t]
\begin{center}
\includegraphics[height=6cm,width=8cm]{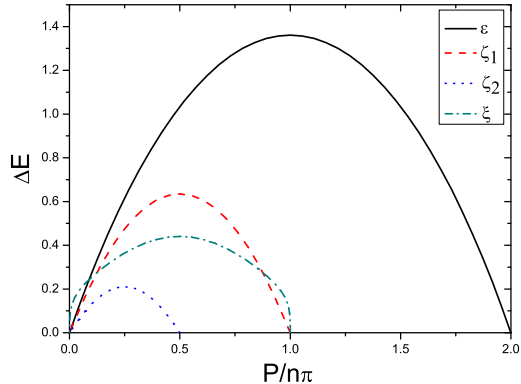}
\end{center}
\caption{(Color online) The single-hole excitations ($Q=1$, $c=1$).
$\Delta E$ and $P$ are the energy and the momentum carried by a
single hole, respectively, and $n$ is the density of particles. The
solid line is the single-hole excitation of real $k$. The dashed
line is that of real $\lambda$. The dotted line is that of $\lambda$
2-string and the dotted dashed line is that of real $\mu$.}
\label{fig3}
\end{figure}
\begin{table}[t]
\caption{\label{table} Possible hole configurations.}
\begin{ruledtabular}
\begin{tabular}{c|cccccc}
$m_1$ &  2&  1&  0&  0&  0&  1\\
$m_2$ &  0&  2&  2&  4&  0&  0\\
$m_3$ &  0&  0&  1&  0&  2&  1
\end{tabular}
\end{ruledtabular}
\end{table}
Nevertheless, the energies of the holes are additive as shown in
Eq.(\ref{de}) and the thermodynamic behavior of the system is mainly
determined by the dispersion relations of the individual holes,
which are shown in Fig.\ref{fig3}. Interestingly, the single
$\lambda$ 2-string hole carries the lowest energy with spin $3/2$
(named as heavy spinon) and therefore dominate the low temperature
thermodynamics of the system.

The simplest spin excitation is a real $\lambda$ hole-pair (the
first column in Table \ref{table}), corresponding to the two domain
walls of a single excited domain. However, unlike the usual spinons,
such holes carry zero spin (named as neutral spinons). The $\lambda$
2-string hole pair can not exist independently but must be
associated with a neutral spinon (the second column in Table
\ref{table}) or a real $\mu$ hole (the third column in Table
\ref{table}). If we add further a $\lambda$ 4-string into a 2-string
hole pair and a real $\lambda$ hole configuration, the total spin of
this excitation is 1. In this case each of the $\lambda$ 2-string
holes carries a spin $1/2$. Such a dressed hole is quite similar to
the ordinary spinon (named as dressed spinon). Four $\lambda$
2-string holes may exist independently (the fourth column in Table
\ref{table}). If we put further one $\lambda$ 6-string and one $\mu$
3-strings into this four hole configuration, we get the $SO(5)$ spin
singlet excitation. The simplest excitation in the $\mu$ sector is a
pair of real $\mu$ holes (the fifth column in Table \ref{table}).
This excitation is quite similar to a real $\lambda$ hole pair but
each of the real $\mu$ hole carries a spin $2$. Joint pair of a real
$\lambda$ hole and a real $\mu$ hole may also happen as shown in the
last column in Table \ref{table}. Other kinds of spin excitations
can be obtained by analyzing the BAE but most of the complex
excitations are generally composed of the excitations listed in
Table \ref{table}.

\acknowledgments

This work was supported by the NSFC, the Knowledge Innovation
Project of CAS, and the National Program for Basic Research of MOST.

* Email: yupeng@aphy.iphy.ac.cn

\end{document}